\documentclass[preprint,3p]{elsarticle}

\usepackage{amssymb}
\usepackage{amsthm}
\usepackage[mathlines]{lineno}
\usepackage{graphicx}
\usepackage{url}
\usepackage{amsmath}
\usepackage{amsfonts}
\usepackage{bm}
\usepackage{textcomp}
\usepackage{subfigure}
\usepackage{multicol}
\usepackage{verbatim}
\usepackage{rotating}
\usepackage[colorlinks,linkcolor=red,citecolor=red]{hyperref}
\usepackage{float}
\usepackage{epsfig}
\usepackage{dcolumn}
\usepackage{bm}
\usepackage{color}
\usepackage{pstricks}
\usepackage{pst-node}
\usepackage{times}
\usepackage{indentfirst}
\usepackage[english]{babel}
\usepackage{siunitx}

\journal{Physics Letters B}

\begin{document}
\begin{frontmatter}

\title{{\bf
\boldmath Cross section measurements of $e^{+}e^{-}\rightarrow p \bar{p} \pi^{0}$ at center-of-mass energies between 4.008 and 4.600 GeV}}

\author{
   \begin{small}
    \begin{center}
      M.~Ablikim$^{1}$, M.~N.~Achasov$^{9,e}$, S. ~Ahmed$^{14}$, X.~C.~Ai$^{1}$, O.~Albayrak$^{5}$, M.~Albrecht$^{4}$, D.~J.~Ambrose$^{44}$, A.~Amoroso$^{49A,49C}$, F.~F.~An$^{1}$, Q.~An$^{46,a}$
% ...
\\
\vspace{0.2cm}
(BESIII Collaboration)\\
\vspace{0.2cm} {\it
$^{1}$ Institute of High Energy Physics, Beijing 100049, People's Republic of China\\
% ...
$^{4}$ Bochum Ruhr-University, D-44780 Bochum, Germany\\
$^{5}$ Carnegie Mellon University, Pittsburgh, Pennsylvania 15213, USA\\
% ...
$^{9}$ G.I. Budker Institute of Nuclear Physics SB RAS (BINP), Novosibirsk 630090, Russia\\
% ...
$^{14}$ Helmholtz Institute Mainz, Johann-Joachim-Becher-Weg 45, D-55099 Mainz, Germany\\
% ...
$^{44}$ University of Rochester, Rochester, New York 14627, USA\\
% ...
$^{46}$ University of Science and Technology of China, Hefei 230026, People's Republic of China\\
% ...
$^{49}$ (A)University of Turin, I-10125, Turin, Italy; (B)University of Eastern Piedmont, I-15121, Alessandria, Italy; (C)INFN, I-10125, Turin, Italy\\
% ...
$^{a}$ Also at State Key Laboratory of Particle Detection and Electronics, Beijing 100049, Hefei 230026, People's Republic of China\\
% ...
$^{e}$ Also at the Novosibirsk State University, Novosibirsk, 630090, Russia\\
% ...
}\end{center}
\vspace{0.4cm}
\end{small}
}

\begin{abstract}
%% Text of abstract
Based on $e^+e^-$ annihilation data samples collected with the BESIII detector at the
BEPCII collider at 13 center of mass energies from 4.008 to 4.600 GeV,
measurements of the Born cross section of $e^{+}e^{-}\rightarrow p \bar{p} \pi^{0}$
are performed. No significant resonant structure is observed in the
measured energy-dependent cross section. The upper limit on the Born
cross section of $e^{+}e^{-}\rightarrow Y(4260) \rightarrow p \bar{p} \pi^{0}$
at the 90\% C.L. is determined to be 0.01 pb.

\text{Keywords: hadrons, cross section measurements, $Y(4260)$}
\end{abstract}

\end{frontmatter}

%\linenumbers
\begin{multicols}{2}
\section{Introduction}
\label{}
 The Born cross section of $e^+e^-\rightarrow p\bar{p}\pi^0$ in
 the vicinity of the $\psi(3770)$ has been measured recently by BESIII~\cite{ppbarpi0_3770}.
 Information on the cross section of $e^+e^-\rightarrow p\bar{p}\pi^0$
 at higher energies is however still lacking.
 The experimental data on the cross section of $e^+e^-\rightarrow hadrons$ can
 be used as an input to calculate the hadronic vacuum polarization via dispersion
 integrals~\cite{eidelman,Davier,Davier2,Hagiwara}.

The charmonium-like state $Y(4260)$ was first observed in its decay to
$\pi^+\pi^-J/\psi$~\cite{Y4260_pipiJpsi}.  So far, there is no evidence of
the $Y(4260)$ in the measured open charm decay channels~\cite{openc1,openc2}
and $R$ value scans~\cite{R1,R2,R3,R4,R5,R6,R7}. Many theoretical models
have been proposed to interpret the nature of $Y(4260)$, \emph{e.g.} as a
tetraquark state~\cite{tetraquark}, a $D_1D$ or $D_0 D^*$ hadronic
molecule~\cite{hadronic}, a hybrid charmonium~\cite{hybrid1,hybrid2},
or a baryonium state~\cite{bary}. Searches for new decay modes of the
$Y(4260)$ may provide information that can shed light on the nature of
$Y(4260)$. In particular, the hybrid model~\cite{hybrid1} predicts a
sizable coupling between the $Y(4260)$ and charmless decays.

In this analysis, we report measurements of the cross section of
$e^+e^-\rightarrow p \bar{p}\pi^0$ based on the $e^+e^-$ annihilation
samples collected with the BESIII detector at 13 center of mass energies
in the range $\sqrt{s} = 4.008-4.600$ GeV as shown in Table~\ref{tab:sec}.
Results of the measurements can be used to estimate the cross section
of $p\bar{p}\rightarrow X_{c\bar{c}}\pi^0$, which is of high importance
for the planned PANDA experiment~\cite{panda} at FAIR in Darmstadt, Germany.

\section{BESIII detector and Monte-Carlo simulation}
The BESIII detector~\cite{bibbes3} is a magnetic spectrometer operating
at BEPCII, a double-ring $e^{+}e^{-}$ collider with center-of-mass
energies between 2.0 and 4.6 GeV and a peak luminosity of
$10^{33}$ $\text{cm}^{-2}\text{s}^{-1}$ near the $\psi(3770)$ mass. The cylindrical
core of the BESIII detector consists of a helium-based
main drift chamber (MDC), a plastic scintillator time-of-flight
system (TOF), and a CsI(Tl) electromagnetic
calorimeter (EMC) that are all enclosed in a superconducting
solenoidal magnet providing a 1.0 T magnetic field.
The solenoid is supported by an octagonal flux-return
yoke with resistive plate counter muon identifier modules
interleaved with steel. The acceptance for charged
particles and photons is 93\% of the 4${\pi}$ solid angle, and
the charged-particle momentum resolution is
0.5\% for transverse momenta of 1 GeV/$c$.
The energy resolution for showers in the EMC is 2.5 (5\%) for
1 GeV photons in the barrel (endcaps) region.

A {\footnotesize GEANT}4-based~\cite{bibgeant4} Monte Carlo (MC)
simulation software package is used to optimize the event selection criteria,
estimate backgrounds and determine the detection
efficiency. For each
energy point, we generate 200,000 signal MC events of $e^{+}e^{-} \rightarrow
p\bar{p}\pi^0$ uniformly in phase space. Effects of initial state radiation (ISR) are
simulated with {\footnotesize KKMC}~\cite{bibkkmc}, where the line shape of the
production cross section of $e^{+}e^{-} \rightarrow
p\bar{p}\pi^0$ is taken from results of the measured cross section
iteratively. Effects of final state radiation off charged particles are simulated
with {\footnotesize PHOTOS}~\cite{bibphotos}.

To study possible backgrounds, a MC sample of inclusive $Y(4260)$
decays, equivalent to an integrated luminosity of 825.6 pb$^{-1}$, is
also generated at $\sqrt{s}$ = 4.26 GeV. In these simulations, the
$Y(4260)$ is allowed to decay generically, with the main known decay
channels being generated using {\footnotesize EVTGEN}~\cite{bibevtgen}
with branching fractions set to world average values~\cite{pdg}. The
remaining events associated with charmonium decays are generated with
{\footnotesize LUNDCHARM}~\cite{biblundcharm}, while the continuum hadronic events
are generated with {\footnotesize PYTHIA}~\cite{bibpythia}. QED events
($e^+ e^- \to e^+ e^-$, $\mu^+ \mu^-$, and $\gamma \gamma$) are
generated with {\footnotesize KKMC}~\cite{bibkkmc}. The sources of backgrounds at other
energy points are assumed to be similar.

\section{Event selection}
The final state in this decay is characterized by two
charged tracks and two photons. Two charged tracks with
opposite charge are required. Each track is required to
have its point of closest approach to the beam axis within
10 cm of the interaction point in the beam direction and
within 1 cm of the beam axis in the plane perpendicular to
the beam. The polar angle of the track is required to be
within the region of $|\cos \theta|<$ 0.93.

The time-of-flight and the specific energy loss $dE/dx$ of
a particle measured in the MDC are combined to calculate
particle identification probabilities for pion, kaon, and
proton hypotheses. For each track, the particle type yielding
the largest probability is assigned. In this analysis, one
charged track is required to be identified as a proton and the
other one as an anti-proton.

Photon candidates are reconstructed using clusters of energy deposited in the EMC.
The energy deposited in nearby TOF counters is included in EMC measurements
to improve the reconstruction efficiency and the energy
resolution. Photon candidates are selected by requiring a minimum
energy deposition of 25 MeV in the barrel EMC ($|\cos \theta|<$ 0.8) or 50 MeV
in the end cap EMC (0.86 $<|\cos \theta|<$ 0.92). To reject photons
radiated from charged particles, the angle between the photon
candidate and the proton is required to be greater than 10 degrees.
A more stringent cut of 30 degrees between the photon candidate
and the anti-proton is applied to exclude the large number of
photons from anti-proton annihilation.

For events with one proton, one anti-proton, and at least
two photons, a kinematic fit (4C) with the total four momenta
of all particles constrained to the energy and
three momentum-components of the initial $e^+e^-$ system is applied.
When more than two photons are found in an
event, all possible $p\bar{p}\gamma\gamma$ combinations are considered
and the one yielding the smallest $\chi^2_{4C}$ is retained for further
analysis. The $\chi^2_{4C}$ is required to be less than 30.
After selecting the $p\bar{p}\gamma\gamma$ candidate, the $\pi^0$ candidates
are selected by requiring $|M(\gamma\gamma) -  m_{\pi^{0}}| < 15$
MeV/$c^{2}$, where $m_{\pi^0}$ is the nominal $\pi^0$ mass~\cite{pdg}.

\begin{figure}[H]
   \centering
   %\subfigure[]{
     \includegraphics[width=0.22\textwidth,height=0.14\textheight]{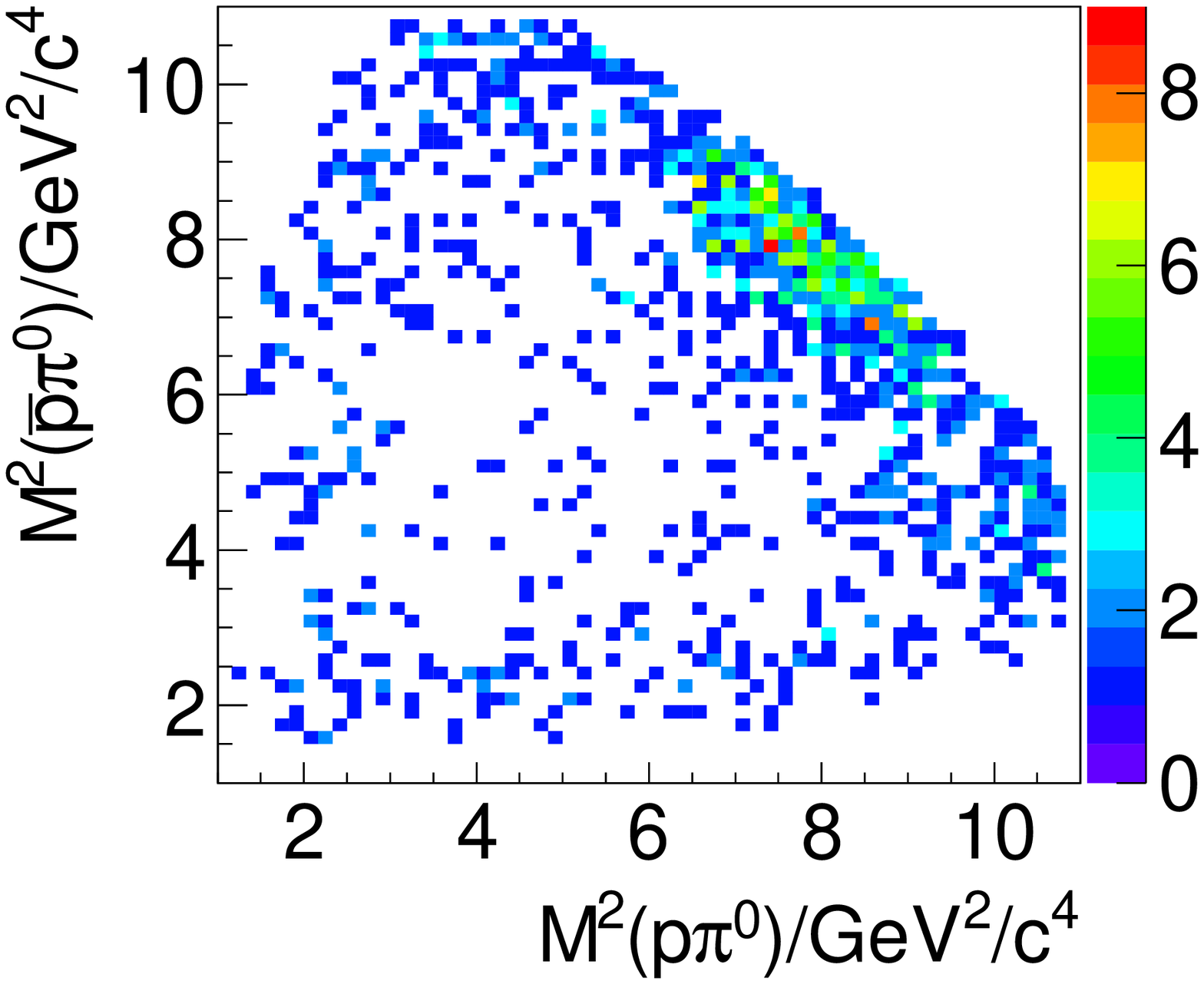}
      \put(-28,75){(a)}
   %\subfigure[]{
     \includegraphics[width=0.22\textwidth,height=0.14\textheight]{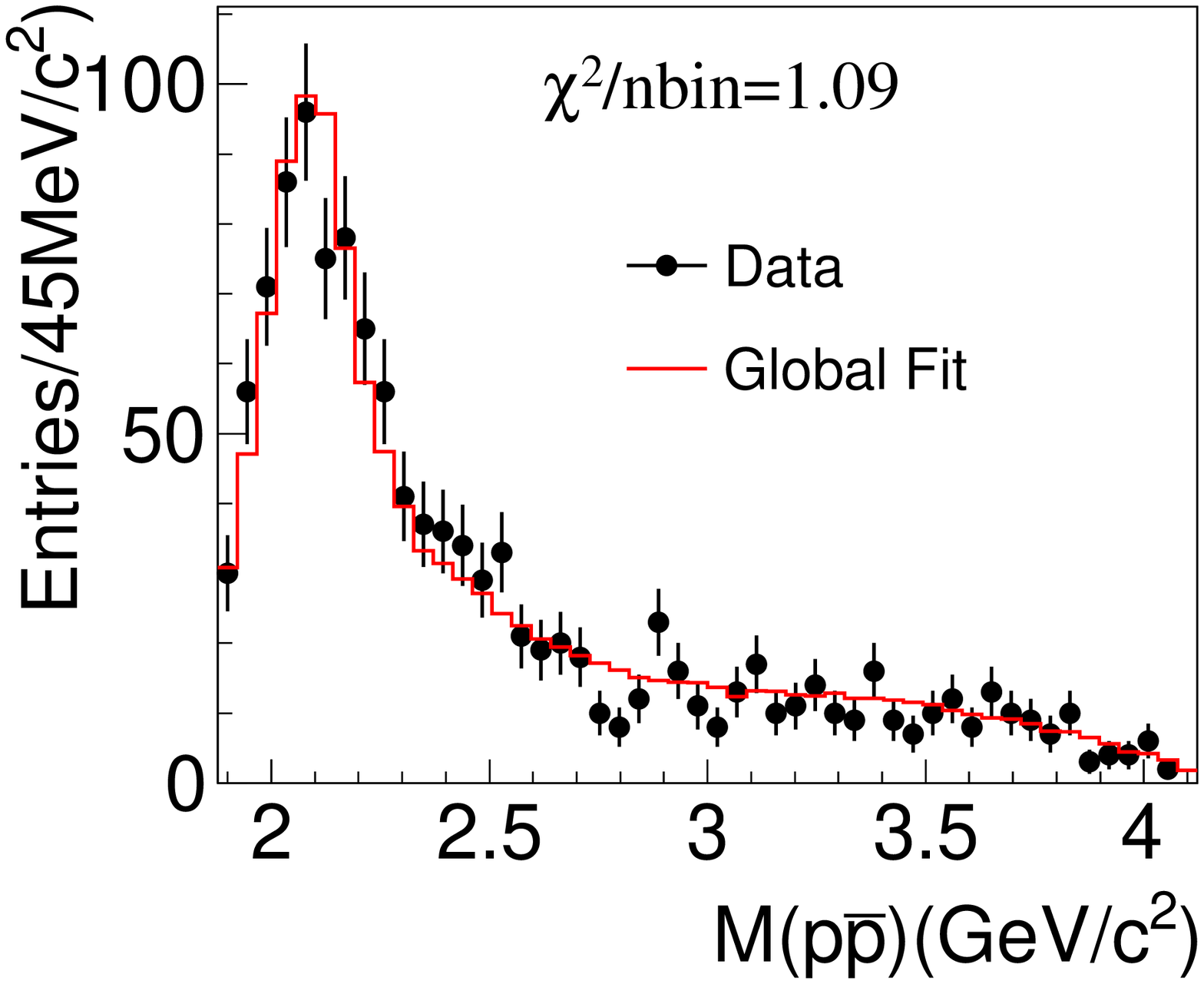}
      \put(-18,75){(b)}\\
   %\subfigure[]{
     \includegraphics[width=0.22\textwidth,height=0.14\textheight]{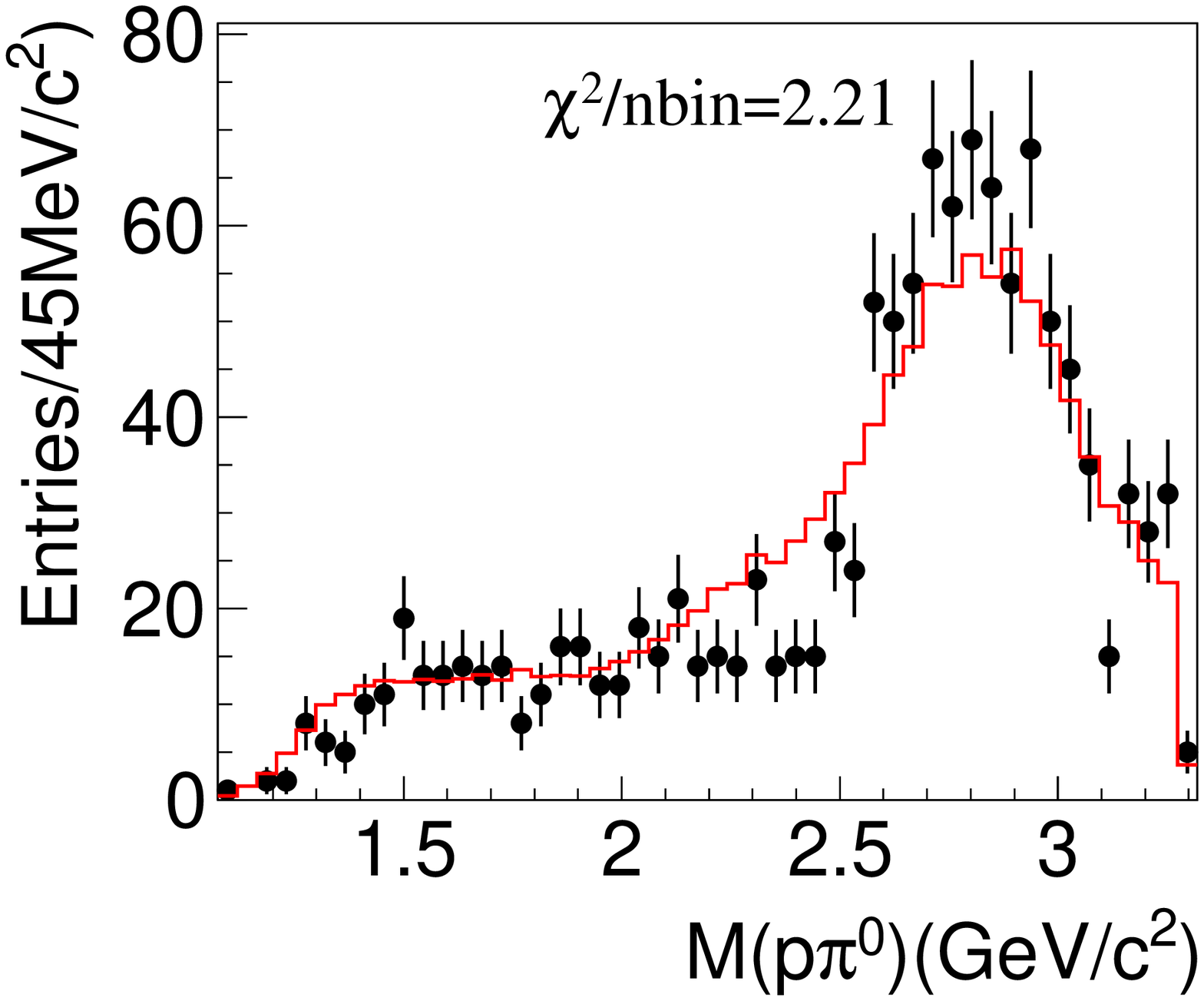}
      \put(-18,75){(c)}
   %\subfigure[]{
     \includegraphics[width=0.22\textwidth,height=0.14\textheight]{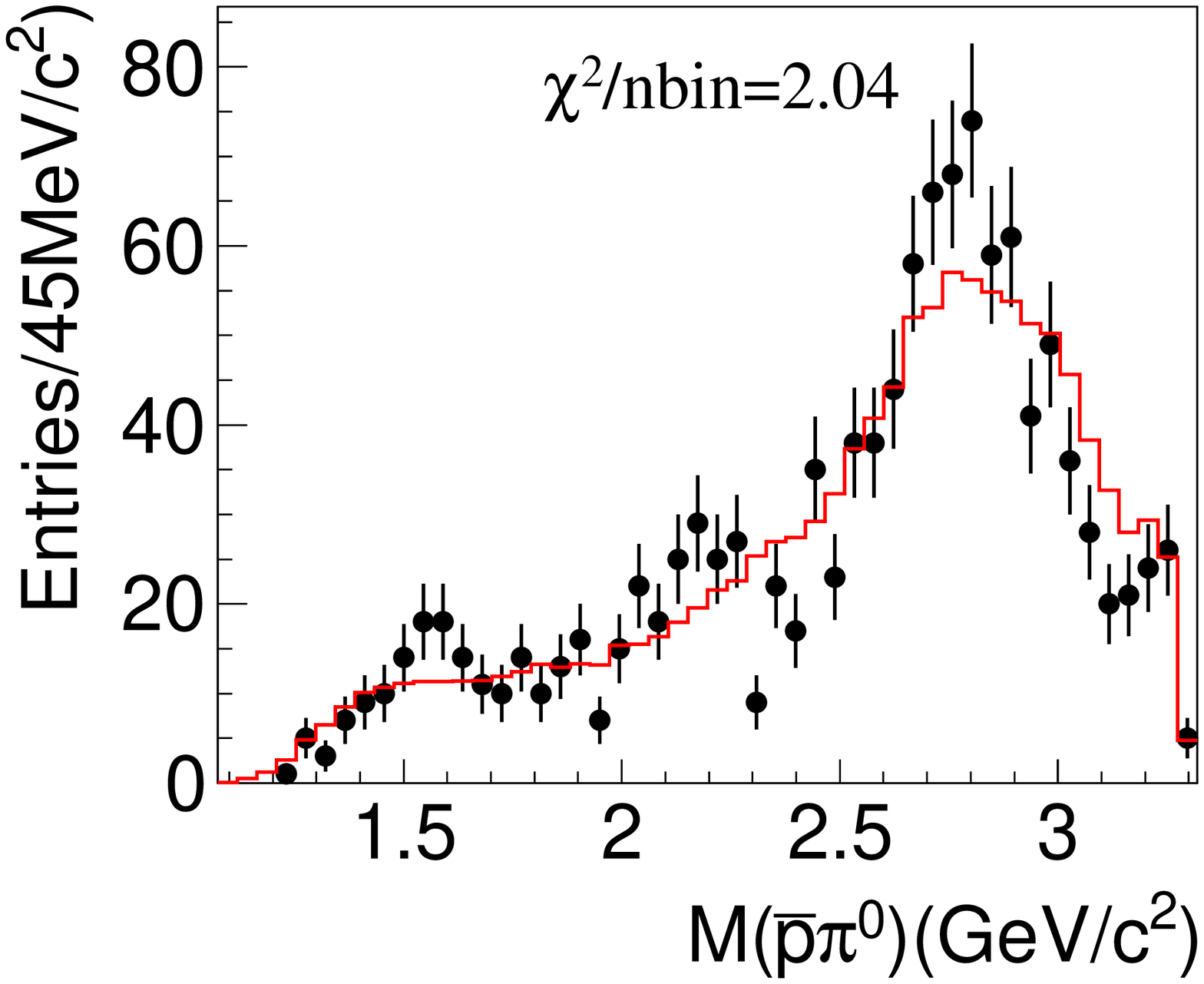}
     \put(-18,75){(d)}
     \caption{(a) Dalitz plot for the selected $e^+e^-\rightarrow
       p\bar{p}\pi^{0}$ candidates of data and invariant mass spectra
       of (b) $p\bar{p}$, (c) $p\pi^0$ and (d) $\bar{p}\pi^0$ at
       $\sqrt{s}$ = 4.258 GeV. In (b), (c) and (d), the points with
       error bars show data and the red histograms show MC projections of
       partial wave analysis fit described in the text.}
     \label{fig:spectra}
 \end{figure}

The Dalitz plot for the events passing the above selection criteria for
data at $\sqrt{s}$ = 4.258 GeV is shown in Fig.~\ref{fig:spectra}(a).
The corresponding invariant mass spectra of $p\bar{p}$, $p\pi^0$ and $\bar{p}\pi^0$
are shown in Fig.~\ref{fig:spectra}(b), (c) and (d), respectively.

 The potential backgrounds for $e^+e^-\rightarrow p\bar{p}\pi^0$ are studied
 using the inclusive MC sample at $\sqrt{s}$ = 4.26 GeV. After imposing all
 event selection requirements, the remaining background events are found
 to have the final state topologies $e^+e^-\rightarrow \gamma p \bar{p}$,
 $\gamma \gamma p \bar{p}$ and $\gamma \gamma \gamma p \bar{p}$.
 No other background survives. The non-$\pi^0$
 background events can be evaluated from events in the $\pi^0$ sidebands. The $\pi^0$
 sideband regions are defined as $0.07 < M(\gamma\gamma) < 0.10$ GeV/$c^2$
 and $0.17 < M(\gamma\gamma) < 0.20$ GeV/$c^2$. The background contamination
 estimated using $\pi^0$ sidebands at $\sqrt{s}$ = 4.258 GeV is 0.3\%. The
 background contributions are neglected in the subsequent analysis.

 \section{Study of intermediate structures by Partial Wave Analysis}
 As shown in Fig.~\ref{fig:spectra}, a prominent structure near the threshold
 in the $p \bar p$ mass spectrum is visible.
 Structures are also seen in the $p\pi^0$ and $\bar p \pi^0$ mass spectra.
 To evaluate the detection efficiencies of the decay $e^+e^-\rightarrow p \bar{p}\pi^0$
 properly, a partial wave analysis (PWA) is performed with the
 $e^+e^-\rightarrow p \bar{p}\pi^0$ candidates to study the intermediate states present.

For the process $e^+e^-\rightarrow p \bar p \pi^0$, the isospin of the $p \bar p \pi^0$
system can be $I=0$ or $I=1$. The quasi-two-body decay amplitudes in the sequential
decay processes $e^+e^-\rightarrow p\bar{N}^*(\bar{p}N^*)$, $N^*(\bar{N}^*)\rightarrow p \pi^0(\bar{p}\pi^0)$, $e^+e^-\rightarrow p\bar{\Delta}^*(\bar{p}\Delta^*)$, $\Delta^*(\bar{\Delta}^*)\rightarrow p \pi^0(\bar{p}\pi^0)$, $e^+e^-\rightarrow \rho^*(\omega^*) \pi^0$, $\rho^*(\omega^*)\rightarrow p\bar{p}$ are constructed in the covariant tensor
formalism~\cite{NBW, cov3}. All $1^{--}$ and $3^{--}$ states above $p\bar{p}$ threshold,
$N^*$ and $\Delta^*$ states with spin up to 5/2, listed in the summary tables of the PDG~\cite{pdg},
are considered in this analysis. According to the framework of soft $\pi$ meson theory~\cite{soft_pi},
the off-shell decay process should be included. Thus, $N(940)$ with a mass of 940 MeV/$c^2$ and zero
width representing a virtual proton which could emit a $\pi^0$ is considered as a possible component.
No isoscalar vector meson is considered, since there is no candidate above the $p \bar p$ threshold
in the summary tables of the PDG. The $\rho^*$ states are parameterized by a constant-width relativistic
Breit-Wigner (BW) propagator with barrier factors included. The $N^*$ and $\Delta^*$ states are
parameterized by a BW propagator as described in Ref.~\cite{NBW}. The resonance parameters are fixed
according to previous measurements~\cite{pdg} due to limited statistics. The complex coefficients of
the amplitudes are determined by an unbinned maximum likelihood fit. The details of the likelihood
function construction can be found in Ref.~\cite{likelihood}.

For $\rho^*$ states with $J=1$, the $p \bar p$ final state interaction (FSI) effect using
the J\"{u}lich~model~\cite{FSI} is taken into consideration by factorizing the partial
wave amplitude into the amplitude without the FSI effect and the $S$ wave $p\bar p$
scattering amplitude in the scattering length approximation given in Ref.~\cite{FSI}. The
direct process of $e^+e^-\rightarrow p \bar{p} \pi^0$ can be modeled by $1^{--}$ or
$3^{--}$ phase space of the $p\bar{p}$ system ($1^{--}$ or $3^{--}$ PHSP). All
combinations of the components in Ref.~\cite{candidates} are evaluated. The changes in the
negative log-likelihood (NLL) and the number of free parameters in the fit with and
without a resonance are used to evaluate its statistical significance. Resonances with
significance greater than 5$\sigma$ are retained in the PWA solution. The selection of PWA
components is performed at the energy points with the high statistics, i.e. at $\sqrt{s}$
= 4.008, 4.226, 4.258 and 4.416 GeV, as shown in Table~\ref{tab:sec}. The selected
components are used to describe the data at other nearby energy points. The data at
$\sqrt{s}= 4.189-4.600$ GeV can be described by the $N(1440)$, $\rho(2150)$,
$\rho_3(1990)$ and $1^{--}$ PHSP amplitudes. The data at $\sqrt{s} = 4.008-4.085$ GeV can
be described by the $N(1520)$, $N(2570)$, $\rho(2150)$, $\rho_3(1990)$ and $1^{--}$ PHSP
amplitudes.  The $N(940)$ is not included in the fits since its significance is less than
5$\sigma$.  If we perform an alternative PWA fit with $N(1440)$, $\rho(2150)$,
$\rho_3(1990)$ and $1^{--}$ PHSP at $\sqrt{s}$ = 4.008 GeV, the NLL worsens by 37.8. The
change of efficiency determined with the alternative fit with respect to the nominal value
is considered as a source of systematic uncertainty.  Comparisons of the data and the fit
projection (weighted by MC efficiencies) in terms of the invariant mass spectra of
$p\bar{p}$, $p\pi^0$ and $\bar{p}\pi^0$ at $\sqrt{s}$ = 4.258 GeV are shown in
Fig.~\ref{fig:spectra}(b), (c) and (d), respectively. The $\chi^2$ over the number of bins
is displayed in those figures.

\section{Cross section for $e^{+}e^{-}\rightarrow p\bar{p}\pi^0$}
The Born cross section for $e^{+}e^{-}\rightarrow p\bar{p}\pi^0$ is determined as
\begin{equation}
\sigma^{B} = \frac{N^\text{obs}}{{\cal L}\cdot (1+\delta^r)\cdot (1+\delta^v)\cdot \epsilon \cdot {\cal B}_{\pi^0}},
\label{eq:sec}
\end{equation}
where $N^\text{obs}$ is the number of observed events; ${\cal L}$ is the integrated
luminosity; $\epsilon$ is the detection efficiency derived from MC events generated
according to the results of the PWA fit; $(1+\delta^r)$ is the radiative correction
factor, which is taken from a QED calculation taking the line shape of the cross section
$e^{+}e^{-}\rightarrow p\bar{p}\pi^0$ of data as input in an iterative procedure; $(1+\delta^v)$ is the
vacuum polarization factor, including leptonic and hadronic contributions, taken from a QED
calculation with an accuracy of 0.5\%~\cite{VP}; and ${\cal B}_{\pi^0}$ is the branching
fraction of $\pi^0$ decaying to $\gamma\gamma$ according to the PDG~\cite{pdg}.  The
measured Born cross section of $e^{+}e^{-}\rightarrow p\bar{p}\pi^0$ at each energy point
is listed in Table~\ref{tab:sec}.

 \begin{table*}[!htbp]
    \centering
     \caption{The results on $e^{+}e^{-}\rightarrow p\bar{p}\pi^0$. Shown in the table are the integrated luminosity ${\cal L}$, the radiative correction factor $(1+\delta^r)$, the vacuum polarization factor $(1+\delta^v)$, the number of observed events $N^\text{obs}$, the detection efficiency $\epsilon$ and the Born cross section $\sigma^{B}(e^{+}e^{-}\rightarrow p\bar{p}\pi^0)$ at each energy point. The errors of $\epsilon$ are from the PWA fit. The first errors of $\sigma^{B}$ are statistical, and the second ones are systematic.}
      \label{tab:sec}
     \begin{small}\sisetup{separate-uncertainty=true}
     \begin{tabular}{cSccS[table-format=3.0,table-figures-uncertainty=2]S[table-format=2.1,table-figures-uncertainty=1]l}
     \hline
     $\sqrt{s}$ (GeV)    &${\cal L}\; [\text{pb}^{-1}]$  &$(1+\delta^r)$  &$(1+\delta^v)$
       &$N^\text{obs}$ &$\epsilon [\%]$ &\multicolumn{1}{c}{$\sigma^{B} [\text{pb}]$}\\
     \hline
     4.008  &482.0  &0.967  &1.044  &1074\pm33  &43.9\pm0.9 &$5.09\pm0.18_{-0.24}^{+0.26}$\\
     4.085  &52.6    &0.992  &1.052 &106\pm11   &43.7\pm1.4 &$4.47\pm0.46_{-0.21}^{+0.27}$\\
     4.189  &43.1    &1.025  &1.056 &75\pm9     &44.7\pm1.0 &$3.64\pm0.43_{-0.19}^{+0.18}$\\
     4.208  &54.6    &1.031  &1.057 &93\pm10    &44.9\pm1.6 &$3.52\pm0.39_{-0.22}^{+0.17}$\\
     4.217  &54.1    &1.034  &1.057 &82\pm10    &43.4\pm1.3 &$3.24\pm0.37\pm0.18$\\
     4.226  &1047.3  &1.037  &1.056 &1611\pm41  &45.2\pm0.5 &$3.15\pm0.08\pm0.14$\\
     4.242  &55.6    &1.042  &1.056 &89\pm9     &44.6\pm1.1 &$3.30\pm0.36_{-0.15}^{+0.19}$\\
     4.258 &825.6    &1.048  &1.054 &1203\pm35  &43.4\pm0.5 &$3.08\pm0.10_{-0.15}^{+0.14}$\\
     4.308 &44.9     &1.063  &1.053 &53\pm8     &46.0\pm1.4 &$2.32\pm0.33_{-0.10}^{+0.15}$\\
     4.358 &539.8    &1.081  &1.051 &668\pm26   &44.7\pm1.1 &$2.48\pm0.11_{-0.12}^{+0.13}$\\
     4.387 &55.2     &1.087  &1.051 &57\pm8     &47.5\pm1.8 &$1.92\pm0.26\pm0.10$\\
     4.416 &1028.9   &1.098  &1.053 &1133\pm34  &44.6\pm0.6 &$2.16\pm0.10_{-0.11}^{+0.10}$\\
     4.600 &566.9    &1.124  &1.055 &474\pm22   &43.8\pm0.8 &$1.63\pm0.08\pm0.08$\\
     \hline
     \end{tabular}
    \end{small}
   \end{table*}

Uncorrelated systematic uncertainties in the Born cross section measurements
mainly originate from the $\pi^0$ mass window requirement, kinematic fit
and the intermediate states in PWA.
The systematic uncertainty from the requirement on the $\pi^0$ signal region
is estimated by smearing the invariant mass of the $\gamma\gamma$ pair in
the signal MC with a Gaussian function to compensate for the resolution difference
between data and MC. The parameters for smearing are determined by fitting the
$\pi^0$ distribution of data with the MC shape convoluted with a Gaussian function.
The difference in the detection efficiency between signal MC samples with and without
the extra smearing is taken as the systematic uncertainty. The systematic uncertainty due
to the kinematic fit is estimated by correcting the helix parameters of charged
tracks for the signal MC sample according the method described in Ref.~\cite{pullcorr}.
The difference in the detection efficiency between the MC samples with and without this
correction is taken as the systematic uncertainty. The systematic uncertainty
from the intermediate states in PWA includes those from the BW parametrization,
resonance parameters and extra resonances. Uncertainties from the BW parametrization
of intermediate states are estimated by replacing the BW formula of $N(1440)$
and $N(1520)$ as used in Ref.~\cite{NBW} with a constant BW formula and replacing
those of $\rho(2150)$ and $\rho_3(1990)$ with the BW formula with the
Gounaris-Sakurai (GS) model~\cite{bibgs}. In the PWA fit, the resonance parameters
are fixed according to the previous measurements~\cite{plb_2150, psip_ppbarpi0}.
Alternative fits are performed in which the resonance parameters are set as free
parameters and the changes in the results are taken as systematic uncertainties.
Uncertainties from additional resonances are estimated by adding the most
significant additional resonance among each $J^{P}$ assignment in Ref.~\cite{candidates}
into the PWA solution individually, and their influences on the cross section
measurements are taken as the systematic uncertainties.

Correlated systematic uncertainties among the different energy points include
those from luminosity measurement (1.0\%)~\cite{luminosity},
MDC tracking (2\% for two charged tracks)~\cite{tracking},
particle identification (2\% in total for proton and anti-proton)~\cite{pid},
photon detection efficiency (2\%)~\cite{photon_sys} and radiative correction.
The difference in $\epsilon(1+\delta^r)$ between the third and fourth iteration
is taken as the systematic uncertainty due to the radiative correction,
as the radiative correction dependent quantity $\epsilon(1+\delta^r)$
converges after three iterations.

The total systematic uncertainty of the different energy
points is calculated by adding the individual uncertainties
in quadrature as shown in Table~\ref{tab:sys_tot}.

 \begin{table*}[!htbp]
\centering
\linespread{1.5}
\caption{Summary of systematic uncertainties on the Born cross section of $e^{+}e^{-}\rightarrow p \bar{p} \pi^{0}$ (\%).}
\label{tab:sys_tot}
\resizebox{1.0\textwidth}{!}{%
\begin{tabular}{cccccccccccccc}
\hline
Sources /$\sqrt{s}$ (GeV)     &4.008 &4.085 &4.189 &4.208 &4.217 &4.226 &4.242 &4.258 &4.308 &4.358 &4.387 &4.416 &4.600 \\
\hline
Luminosity                     &1.0 &1.0 &1.0 &1.0 &1.0 &1.0 &1.0 &1.0 &1.0 &1.0 &1.0 &1.0 &1.0\\
MDC tracking                   &2.0 &2.0 &2.0 &2.0 &2.0 &2.0 &2.0 &2.0 &2.0 &2.0 &2.0 &2.0 &2.0\\
PID                            &2.0 &2.0 &2.0 &2.0 &2.0 &2.0 &2.0 &2.0 &2.0 &2.0 &2.0 &2.0 &2.0  \\
Photon detection               &2.0 &2.0 &2.0 &2.0 &2.0 &2.0 &2.0 &2.0 &2.0 &2.0 &2.0 &2.0 &2.0  \\
Kinematic fit                  &2.1 &1.9 &1.8 &1.6 &1.7 &1.6 &1.6 &2.1 &1.5 &1.8 &1.5 &1.8 &1.6\\
$\pi^0$ mass resolution        &0.2 &0.2 &0.2 &0.2 &0.2 &0.3 &0.2 &0.4 &0.6 &0.4 &0.3 &0.4 &0.4   \\
Radiative correction           &1.9 &1.9 &1.9 &1.9 &1.9 &1.9 &1.9 &1.9 &1.9 &1.9 &1.9 &1.9 &1.9   \\
Intermediate states in PWA &$_{-1.3}^{+2.4}$ &$_{-1.3}^{+4.1}$ &$_{-2.6}^{+2.1}$ &$_{-4.4}^{+2.2}$ &$_{-3.5}^{+3.3}$  &$_{-1.1}^{+1.0}$ &$_{-1.0}^{+3.8}$  &$_{-1.9}^{+0.4}$ &$_{-0.9}^{+4.8}$ &$_{-1.7}^{+2.8}$  &$_{-2.3}^{+3.1}$ &$_{-2.1}^{+0.9}$  &$\pm$1.9  \\
\hline
Total  &$_{-4.8}^{+5.2}$ &$_{-4.7}^{+6.1}$ &$_{-5.2}^{+4.9}$ &$_{-6.2}^{+4.9}$ &$_{-5.6}^{+5.5}$  &$\pm4.5$ &$_{-4.5}^{+5.8}$  &$_{-5.0}^{+4.6}$ &$_{-4.5}^{+6.5}$ &$_{-4.8}^{+5.3}$  &$_{-4.9}^{+5.4}$ &$_{-4.9}^{+4.6}$  &$\pm$4.8                            \\
\hline
\end{tabular}
}
\end{table*}

\section{Upper limit on $\sigma(e^+e^-\rightarrow Y(4260)\rightarrow p \bar{p}\pi^0)$}
\label{sect:Y4260}
\begin{figure}[H]
   \centering
   %\subfigure[]{
     \includegraphics[width=0.4\textwidth,height=0.22\textheight]{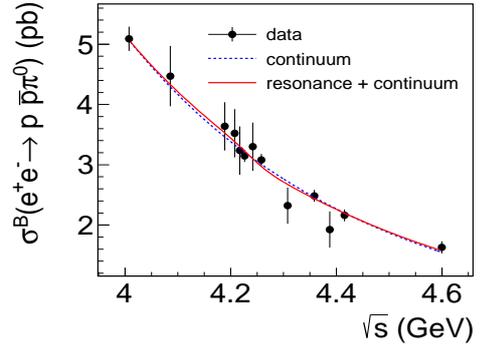}
   %\subfigure[]{
     \caption{Fit to $\sigma(e^+e^-\rightarrow p \bar{p}\pi^0)$ with resonance and continuum (solid line), or only continuum term (dashed line). Dots with error bars are the measured Born cross sections. The uncertainties are statistical only.}
     \label{fig:sec_plot}
  \end{figure}

Figure~\ref{fig:sec_plot} shows the measured Born cross section
of $e^{+}e^{-}\rightarrow p\bar{p}\pi^0$ in the energy region
studied in this work. No significant resonant structure is observed.
The upper limit on the Born cross section of $e^+e^-\rightarrow Y(4260)\rightarrow p \bar{p}\pi^0$
is determined by a least squares fit of
\begin{equation}
\sigma(s) = \left|\sqrt{\sigma_\text{con}} + \sqrt{\sigma_{Y}}\frac{m\Gamma}{s - m^2 + im\Gamma}\exp(i\phi)\right|^2,
\label{eq:sigma_s}
\end{equation}
to the calculated cross sections. In Eq.~(\ref{eq:sigma_s}), $\sigma_\text{con}$ and $\sigma_Y$
represent the continuum cross section and resonant cross section, respectively,
and $\sigma_\text{con}$ can be described by a function of $s$, $\sigma_\text{con}=C/s^\lambda$, where the
exponent $\lambda$ is \emph{a priori} unknown. The parameter $\phi$ describes the phase
between resonant and continuum production amplitudes.
The mass $m$ and width $\Gamma$ of the $Y(4260)$ are fixed to the PDG values~\cite{pdg}.
The values of $C$, $\lambda$, $\sigma_Y$, and the interference phase $\phi$ are free in the fit.
The uncorrelated systematic uncertainties in the Born cross section measurements are
directly considered in the fit and the effect of the correlated systematic uncertainties
on the final results is estimated by the method in Ref.~\cite{offsetmethod}, in which
the error propagation is determined from shifting the data by the aforementioned
correlated uncertainties and adding the deviations in quadrature.
In addition, the uncertainties for the beam energy measurements of all the data points
taken from Ref.~\cite{cme} are considered in the fit.
The best fit function is shown in Fig.~\ref{fig:sec_plot} as the solid line.
The dashed line represents the fit with $\sigma_Y$ = 0. The optimal value of $\sigma_Y$
is $(1.6\pm5.9)\times10^{-3}$ pb with a statistical significance of 0.5$\sigma$.
The significance is calculated based on the changes in the $\chi^2$ value and the
number of free parameters in the fit with and without the assumption of existence
of the $Y(4260)$ resonance. The result for the phase between resonant and continuum
production amplitudes is $\phi$ = $3.4\pm1.0$. The parameters describing the slope
of the continuum cross section are $C$ = $(5.4\pm5.3)\cdot10^5$ GeV$^{2\lambda}$pb
and $\lambda$ = $4.2\pm0.4$.
The upper limit on $\sigma_Y$ at the 90\% C.L., $\sigma_Y^{up}$, is determined
by $\int_{0}^{\sigma_Y^{up}} G(\sigma_Y, \sigma_{\sigma_Y})dx/\int_{0}^{\infty}G(\sigma_Y, \sigma_{\sigma_Y})dx=0.9$,
where $G(\sigma_Y, \sigma_{\sigma_Y})$ is a Gaussian function with mean value $\sigma_Y = 1.6\times 10^{-3}$ pb
and standard deviation $\sigma_{\sigma_Y} = 5.9\times 10^{-3}$ pb. The uncertainties
from mass and width of the $Y(4260)$ are considered by varying them by one standard
deviation according to the PDG values~\cite{pdg} and the most conservative $\sigma_Y^{up}$
is taken as the final result. The obtained upper limit is 0.01 pb.

\section{Summary}
Based on 13 data samples between $\sqrt{s}$ = 4.008 and 4.600 GeV collected with
the BESIII detector, the process $e^+e^-\rightarrow p\bar{p}\pi^0$ is studied.
The Born cross section of $e^+e^-\rightarrow p\bar{p}\pi^0$ is measured.
No resonant structure is observed in the shape of the cross section.
The upper limit on the Born cross section of $e^+e^-\rightarrow Y(4260)\rightarrow p \bar{p}\pi^0$
at the 90\% C.L is estimated to be 0.01 pb.

\section{Acknowledgements}
The BESIII collaboration thanks the staff of BEPCII and the IHEP computing center for their strong support. This work is supported in part by National Key Basic Research Program of China under Contract No. 2015CB856700; National Natural Science Foundation of China (NSFC) under Contracts Nos. 11235011, 11305178, 11322544, 11335008, 11375204, 11425524, 11635010; the Chinese Academy of Sciences (CAS) Large-Scale Scientific Facility Program; the CAS Center for Excellence in Particle Physics (CCEPP); the Collaborative Innovation Center for Particles and Interactions (CICPI); Joint Large-Scale Scientific Facility Funds of the NSFC and CAS under Contracts Nos. U1232201, U1332201, U1532257, U1532258; CAS under Contracts Nos. KJCX2-YW-N29, KJCX2-YW-N45; 100 Talents Program of CAS; National 1000 Talents Program of China; INPAC and Shanghai Key Laboratory for Particle Physics and Cosmology; German Research Foundation DFG under Contracts Nos. Collaborative Research Center CRC 1044, FOR 2359; Istituto Nazionale di Fisica Nucleare, Italy; Koninklijke Nederlandse Akademie van Wetenschappen (KNAW) under Contract No. 530-4CDP03; Ministry of Development of Turkey under Contract No. DPT2006K-120470; The Swedish Resarch Council; U. S. Department of Energy under Contracts Nos. DE-FG02-05ER41374, DE-SC-0010504, DE-SC0010118, DE-SC0012069; U.S. National Science Foundation; University of Groningen (RuG) and the Helmholtzzentrum fuer Schwerionenforschung GmbH (GSI), Darmstadt; WCU Program of National Research Foundation of Korea under Contract No. R32-2008-000-10155-0.

\end{multicols}

\begin{thebibliography}{**}
%\bibliographystyle{elsarticle-harv}
\bibitem{ppbarpi0_3770} M. Ablikim {\it et al}. (BESIII Collaboration), Phys. Rev. D {\bf 90}, 032007 (2014).
\bibitem{eidelman} S. Eidelman and F. Jegerlehner, Z. Phys. C {\bf 67}, 585 (1995).
\bibitem{Davier} M. Davier, S. Eidelman, A. H\"{o}cker and Z. Zhang, Eur. Phys. J. C {\bf 31}, 503 (2003).
\bibitem{Davier2} M. Davier, A. H\"{o}cker, B. Malaescu and Z. Zhang, Eur. Phys. J. C {\bf 71}, 1515 (2011).
\bibitem{Hagiwara} K. Hagiwara, R. Liao, A.D. Martin, Daisuke Nomura, T. Teubner, J. Phys. G {\bf 38}, 085003 (2011).
\bibitem{Y4260_pipiJpsi}  B. Aubert {\it et al}. (BABAR Collaboration), Phys. Rev. Lett. {\bf 95},
142001 (2005)
\bibitem{openc1} G. Pakhlova {\it et al}. (Belle Collaboration), Phys. Rev. Lett. {\bf 98}, 092001 (2007); Phys. Rev. D {\bf 77},
011103 (2008); Phys. Rev. Lett. {\bf 100}, 062001 (2008); Phys. Rev. D {\bf 80}, 091101 (2009).
\bibitem{openc2} B. Aubert {\it et al}. (BABAR Collaboration), Phys. Rev. D {\bf 79},
092001 (2009).
\bibitem{R1} J. Burmester {\it et al}. (PLUTO Collaboration), Phys. Lett. B {\bf 66},
395 (1977).
\bibitem{R2} R. Brandelik {\it et al}. (DASP Collaboration), Phys. Lett. B {\bf 76},
361 (1978).
\bibitem{R3} J. Siegrist {\it et al}., Phys. Rev. D {\bf 26}, 969 (1982).
\bibitem{R4} A. Osterheld {\it et al}., SLAC Report No. SLAC-PUB-4160, 1986.
\bibitem{R5} J. Z. Bai {\it et al}. (BES Collaboration), Phys. Rev. Lett. {\bf 84}, 594 (2000); Phys. Rev. Lett. {\bf 88}, 101802 (2002).
\bibitem{R6} D. Cronin-Hennessy {\it et al}. (CLEO Collaboration), Phys. Rev. D {\bf 80}, 072001 (2009).
\bibitem{R7} M. Ablikim {\it et al}. (BES Collaboration), Phys. Lett. B {\bf 677}, 239 (2009).
\bibitem{tetraquark} L. Maiani, V. Riquer, F. Piccinini, and A. D. Polosa,
Phys. Rev. D {\bf 72}, 031502 (2005).
\bibitem{hadronic} G. J. Ding, Phys. Rev. D {\bf 79}, 014001 (2009).
\bibitem{hybrid1}  F. E. Close and P. R. Page, Phys. Lett. B {\bf 628}, 215 (2005).
\bibitem{hybrid2} S. L. Zhu, Phys. Lett. B {\bf 625}, 212 (2005).
\bibitem{bary} C. F. Qiao, Phys. Lett. B {\bf 639}, 263 (2006).
\bibitem{panda} M. Lutz {\it et al}.(PANDA Collaboration), arXiv:0903.3905.
\bibitem{bibbes3} M. Ablikim {\it et al}. (BESIII Collaboration), Nucl. Instrum. Methods Phys. Res., Sect. A {\bf 614}, 345 (2010).
\bibitem{bibgeant4} S. Agostinelli {\it et al}. (GEANT4 Collaboration), Nucl.
Instrum. Methods Phys. Res., Sect. A {\bf 506}, 250 (2003).
\bibitem{bibkkmc} S. Jadach, B. F. L. Ward, and Z. Was, Comput. Phys. Commun. {\bf 130}, 260 (2000); Phys. Rev. D {\bf 63}, 113009 (2001).
\bibitem{bibphotos} P. Golonka and Z. Was, Eur. Phys. J. C {\bf 45}, 97 (2006)
\bibitem{bibevtgen} D. J. Lange, Nucl. Instrum. Methods Phys. Res., Sect. A {\bf 462}, 152 (2001) Phys. C {\bf 38}, 090001 (2014).
\bibitem{pdg} K. A. Olive {\it et al}. (Particle Data Group Collaboration), Chin. Phys. C {\bf 38}, 090001 (2014).
\bibitem{biblundcharm} R. G. Ping, Chin. Phys. C {\bf 32}, 599 (2008).
\bibitem{bibpythia} T. Sj\"{o}strand {\it et al}., arXiv:hep-ph/0108264.
\bibitem{NBW} M. Ablikim {\it et al}. (BES Collaboration), Phys. Rev. D {\bf 80}, 052004 (2009).
\bibitem{cov3} W. H. Liang, P. N. Shen, J. X. Wang, and B. S. Zou, J. Phys. G {\bf 28}, 333 (2002).
\bibitem{soft_pi} L. Adler and R. F. Dashen, \emph{Current Algebra and
Application to Particle Physics} (Benjamin, New York, 1968); B. W. Lee, \emph{Chiral Dynamics} (Gordon and Breach,
New York, 1972).
\bibitem{likelihood} M. Ablikim {\it et al}. (BESIII Collaboration), Phys. Rev. D {\bf 87}, 092009 (2013).
\bibitem{FSI} A. Sibirtsev {\it et al}., Phys. Rev. D {\bf 71}, 054010 (2005).
\bibitem{candidates} We considered the following resonance candidates in PDG: $\rho(1700)$, $\rho(1900)$, $\rho(2150)$, $\rho(2000)$, $\rho(2270)$, $\rho_3(1690)$, $\rho_3(1990)$, $\rho_3(2250)$, $N(1535)$, $N(1650)$, $N(1710)$, $N(1440)$, $N(2100)$, $N(2300)$, $N(1520)$, $N(1720)$, $N(1675)$, $N(2060)$, $N(2570)$, $N(1680)$, $N(940)$, $\Delta(1620)$, $\Delta(1910)$, $\Delta(1700)$, $\Delta(1232)$, $\Delta(1600)$, $\Delta(1920)$, $\Delta(1930)$, $\Delta(1905)$, $1^{--}$ PHSP and $3^{--}$ PHSP.
\bibitem{VP} S. Actis {\it et al.}, Eur. Phys. J. C {\bf 66}, 585 (2010).
\bibitem{pullcorr} M. Ablikim {\it et al}. (BESIII Collaboration), Phys. Rev. D {\bf 87}, 012002 (2013).
\bibitem{bibgs} J. P. Lees {\it et al}. (BABAR Collaboration), Phys. Rev. D {\bf 86}, 032013 (2012).
\bibitem{plb_2150} A. V. Anisovich {\it et al.}, Phys. Lett. B {\bf 491}, 47 (2000); Phys. Lett. B {\bf 508}, 6 (2001); Phys. Lett. B {\bf 513}, 281 (2001); Phys. Lett. B {\bf 542}, 8 (2002).
\bibitem{psip_ppbarpi0} M. Ablikim {\it et al}. (BESIII Collaboration), Phys. Rev. Lett. {\bf 110}, 022001 (2013).
\bibitem{luminosity} M. Ablikim {\it et al}. (BESIII Collaboration), Chin. Phys. C {\bf 39}, 093001 (2015).
\bibitem{tracking}  W.~L. Yuan {\it et al.}, arXiv:1507.03453 [hep-ex].
\bibitem{pid} M. Ablikim {\it et al}. (BESIII Collaboration), Phys. Rev. D {\bf 86}, 032014 (2012).
\bibitem{photon_sys} M. Ablikim {\it et al}. (BESIII Collaboration), Phys. Rev. D {\bf 83}, 112005 (2011).
\bibitem{offsetmethod} M. Botje, J. Phys. G {\bf 28}, 779 (2002).
\bibitem{cme} M. Ablikim {\it et al}. (BESIII Collaboration), Chin. Phys. C {\bf 40}, 063001 (2016).
\end{thebibliography}
\end{document}